\documentclass[journal=nalefd,manuscript=article,layout=twocolumn]{achemso}

\usepackage[version=3]{mhchem} 
\usepackage{siunitx}
\usepackage{parskip}
\usepackage{amssymb}
\usepackage{ulem}
\usepackage{lineno}
\usepackage{subcaption}
\usepackage{graphicx}
\DeclareSIUnit\angstrom{\protect \text {Å}}

\author{Martin Werres}
\affiliation{Institute of Engineering Thermodynamics, German Aerospace Center (DLR), Wilhelm-Runge-Str. 10, 89081 Ulm, Germany}
\alsoaffiliation[HIU]{Helmholtz Institute Ulm (HIU), Helmholtzstr. 11, 89081 Ulm, Germany}
\author{Dariusz Niedziela}
\affiliation{Department Flow and Material Simulation, Fraunhofer Institute for Industrial Mathematics, Fraunhofer-Platz 1, D-67663 Kaiserslautern, Germany}

\author{Arnulf Latz}
\affiliation{Institute of Engineering Thermodynamics, German Aerospace Center (DLR), Wilhelm-Runge-Str. 10, 89081 Ulm, Germany}
\alsoaffiliation[HIU]{Helmholtz Institute Ulm (HIU), Helmholtzstr. 11, 89081 Ulm, Germany}
\alsoaffiliation[University Ulm]{Department of Electrochemistry, University of Ulm, Albert-Einstein-Allee 47, 89081 Ulm, Germany}
\author{Birger Horstmann}
\affiliation{Institute of Engineering Thermodynamics, German Aerospace Center (DLR), Wilhelm-Runge-Str. 10, 89081 Ulm, Germany}
\alsoaffiliation[HIU]{Helmholtz Institute Ulm (HIU), Helmholtzstr. 11, 89081 Ulm, Germany}
\alsoaffiliation[University Ulm]{Department of Electrochemistry, University of Ulm, Albert-Einstein-Allee 47, 89081 Ulm, Germany}
\email{birger.horstmann@dlr.de}

\let\oldmaketitle\maketitle
\let\maketitle\relax

\title
  {Stress-driven whisker formation in lithium metal batteries}

\keywords{Lithium metal battery, Lithium whiskers, Simulation, SEI, Fluid dynamics}

\begin{document}

\twocolumn[
\begin{@twocolumnfalse}
\oldmaketitle
\begin{abstract}
Lithium metal batteries are promising for next-generation high-energy-density batteries, especially when lithium is directly plated on a current collector. However, lithium whiskers can form in the early stages of electroplating. These whiskers lead to low Coulombic efficiency due to isolated lithium formation during stripping. The mechanism of whisker formation is not fully understood, and different mechanisms are proposed in the literature. Herein, we computationally explore a stress-driven extrusion mechanism through cracks in the solid-electrolyte-interphase (SEI), which explains the experimentally observed root growth of lithium whiskers. We model the extrusion as a flow of a power-law Herschel-Bulkley fluid parametrized by the experimental power-law creep behavior of lithium, which results in the typical one-dimensional whisker shape. Consequently, in competition with SEI self-healing, SEI cracking determines the emergence of whiskers, giving a simple rule of thumb to avoid whisker formation in liquid electrolytes.
\end{abstract}
\end{@twocolumnfalse}
]

\clearpage
\footnotesize  
\section{Introduction}

Lithium metal anodes paired with liquid electrolytes are promising candidates for next-generation high-energy-density batteries\cite{Lin2017,Liu2018,Ren2019,Horstmann2021}. Energy density is highest when lithium is directly plated on a current collector\cite{Qian2016,Tian2020}. However, this setup requires ultra-high Coulombic Efficiency (CE) because lithium inventory loss is not absorbed by excess lithium\cite{Qian2016}. CE can be lowered by (1) side reactions that form solid-electrolyte interphase (SEI) and (2) electronic disconnection of lithium from the current collector during stripping, called \textit{isolated lithium}\cite{Fang2019,Hobold2021}. Conventional carbonate-based electrolytes have low CE because of \textit{lithium whisker} formation during plating\cite{Yoshimatsu1988,Steiger2014a,Xu2020,Becherer2022}. This morphology is undesirable, as the whiskers grow into mossy lithium\cite{Bai2016}, which enhances SEI formation, increases cell resistance\cite{Chen2017}, and leads to isolated lithium formation during stripping\cite{Yoshimatsu1988,Steiger2014b,Chen2017,Werres2023}. Here, we computationally study the early whisker formation of lithium electroplated on copper to build a fundamental understanding of lithium whisker formation, which can help mitigate the problem.


Lithium whiskers are needle-like lithium deposits, which typically have a small diameter in the order of $\SI{100}{nm}$ to $\SI{1}{\mu m}$ and are several micrometers in length\cite{Yamaki1998,Xu2020}. Whiskers are single-crystals, but crystal orientation may change with whisker kinks, where the deposition suddenly changes growth direction\cite{Li2017a}. These observations have been confirmed by cryogenic electron microscopy, which can resolve the small length scale while minimizing beam damage\cite{Li2017a,Li2018,Zachman2018,Xu2020,Xu2020_2,Huang2020}. The term \textit{whisker} does not imply a specific growth mechanism but is useful for verbally distinguishing the structures from lithium dendrites. Lithium dendrites grow from transport limitations in the electrolyte when the salt concentration at the anode approaches zero after Sand's time. Lithium whiskers are observed when dendrite conditions are not met and thus have a different growth mechanism\cite{Bai2016}. 


Several mechanisms for lithium whisker growth have been discussed in the literature. Electromigration was already ruled out to play a significant role in the growth of lithium whiskers\cite{Tang2016,Rulev2019}. Experiments showed that lithium whiskers have no clear growth direction\cite{Steiger2014a,Kushima2017,Becherer2022}. Secondly, the intrinsic metal properties, such as the surface self-diffusion, are considered to explain heterogeneous plating\cite{Jckle2014}. While this can explain the tendency of lithium to grow rough surfaces, it does not account for the fact that lithium morphology depends on the SEI\cite{Horstmann2021}. Steiger \textit{et al.} observed that whiskers grow by atom insertion at the base or whisker kinks\cite{Steiger2014a}. Observations from \textit{in situ} microscopy experiments of lithium whisker growth during electroplating suggest that lithium whiskers initially grow from the root\cite{Yamaki1998,Steiger2014a,Kushima2017,Yulaev2018}. The most promising mechanism for understanding the root-growth behavior is a stress-driven mechanism: the built-up compressive stress in lithium underneath the SEI is released through cracks in the SEI\cite{Yamaki1998,Tang2016,Kushima2017,Wang2018}. The mechanism is known from the formation of tin whiskers in lead-free soldering\cite{Ills2017,Majumdar2019}. This mechanism captures the root-growth mode of lithium whiskers and the insignificant effect of electric field lines in the vicinity of the whiskers. We, therefore, explore this mechanism in our model to verify whether the conditions in the early stages of electroplating can describe the onset of lithium whisker growth. 

In tin whisker research, there is a consensus that three conditions are necessary to explain whisker growth: (1) in-plane compressive stress to provide the driving force of whisker growth, (2) a matter transport mechanism to the whisker, \textit{e.g.}, rapid grain boundary self-diffusion, and (3) a surface layer limiting stress relief via diffusional processes\cite{Majumdar2019}. For lithium plating, the stress can be plating induced\cite{Wang2018}. The induced stress can lead to creep under battery conditions, explaining the fast matter transport. Experiments showed that lithium experiences power-law creep with a stress exponent of about 6.6, indicating that dislocation creep is the dominant mechanism\cite{Masias2018,LePage2019}. Finally, lithium is inherently passivated by the SEI, which limits stress relief. 

The stability of the SEI passivation layer is essential to understanding the emergence of whiskers, as shown by vacuum experiments, where only a native Li$_2$O passivation layer forms due to trace amounts of oxygen\cite{Yulaev2018}. While few trace amounts of oxygen (no stable surface layer) and large quantities of trace amounts (thick and stable surface layer) lead to the formation of microparticles, an intermediate trace amount of oxygen leads to whisker formation. There, the surface layer was thick enough to hinder surface diffusion but not stable enough to prevent lithium whisker root growth once the SEI cracks\cite{Yulaev2018}. 

We study the emergence of lithium whiskers during electroplating on copper in liquid electrolyte. We first rationalize the geometry of whiskers and the conditions under which whisker formation is expected. The cracking of the SEI is a necessary condition for root-driven whisker growth. Therefore, we discuss the relationship between the SEI properties and operating conditions with the cracking of the SEI. We model a lithium nucleus covered by an SEI with a pre-existing crack and a steady lithium influx due to the plating conditions. The SEI crack is a weak spot where lithium can extrude into the electrolyte. The flow of lithium is modeled by a power-law Herschel-Bulkley fluid derived from lithium creep observations. Our simulation results suggest that the experimentally observed whisker shape is induced by extrusion. Plating on copper with carbonate-based electrolytes favors whisker formation because, in the early stage after nucleation, the conditions for whisker growth are met. We expect the presented mechanism to occur on various electrodes with heterogeneities, \textit{e.g.}, lithium foils with high surface roughness, lithium foils under high overpotentials initiating nucleation, and lithium plated on graphite anodes.

\section{Theory}

Firstly, we want to argue why lithium whiskers are observed with the typical diameter of $\SI{100}{nm}$ to $\SI{1}{\mu m}$. Experimentally, Pei et al. studied the lithium nucleation behavior on copper and observed particle sizes in the order of micrometer\cite{Pei2017}. The SEI growing on the lithium particles experiences significant stresses; it will deform and, in some cases, fracture. Fracture typically occurs for strains in the order of magnitude of $10\%$. The SEI would then relax and leave a hole in the order of $10\%$ of the particle radius behind. Lithium ions continuously pass through the still intact SEI and electroplate beneath it, causing new stress build-up. The extrusion of lithium through the hole in the SEI relieves the new stress and explains the typically observed whisker diameter. This assumption holds if the stress is not too high, \textit{i.e.} so that the SEI does not break further.


Secondly, we want to discuss the conditions for local SEI cracking and whisker emergence. The observation of whiskers in experiments depends on the electrolyte formulation and operating conditions. The main difference when using different electrolytes is the derived SEI. As discussed, the stability of the covering SEI plays a crucial role in understanding the early lithium morphology. To simplify the problem, we consider an existing lithium nucleus with a thin initial SEI. The continuous growth of the nucleus stretches the SEI\cite{vonKolzenberg2021,Kbbing2024}, while the SEI grows in thickness in a diffusion-limited fashion\cite{Peled1979,Single2018,vonKolzenberg2020,Kbbing2023}. This translates to a system of coupled differential equations for the nucleus radius $r_\text{nucleus}$ and the SEI thickness $L_\text{SEI}$:
\begin{equation}
\label{eq:SEI_growth}
    \begin{aligned}
        \frac{\text{d}r_\text{nucleus}}{\text{d}t}&=\frac{I \cdot V_\text{m}^\text{Li}}{2 \pi r_\text{nucleus}^2 F} \\
        \frac{\text{d}L_\text{SEI}}{\text{d}t}&=\frac{V_\text{m}^\text{SEI} D_{e^-}c_{e^-}}{L_\text{SEI}} - \frac{2 L_\text{SEI}}{r_\text{nucleus}}\cdot  \frac{\text{d}r_\text{nucleus}}{\text{d}t} \ .
    \end{aligned}
\end{equation}
Here, $I$ is the current per nucleus, $V_\text{m}^\text{Li}$ is the molar volume of lithium, $F$ is the Faraday constant, $V_\text{m}^\text{SEI}$ is the molar volume of the SEI, $D_{e^-}$ is the diffusivity of electrons in the SEI, and $c_{e^-}$ is the concentration of electrons in the SEI. We evaluate these differential equations with typical parameters\cite{vonKolzenberg2020}, see SI Table~1, to get an estimate about the order of magnitude and when to expect (un)stable SEI conditions. 

\begin{figure}[t]
    \centering  
    \includegraphics[width=0.99\linewidth]{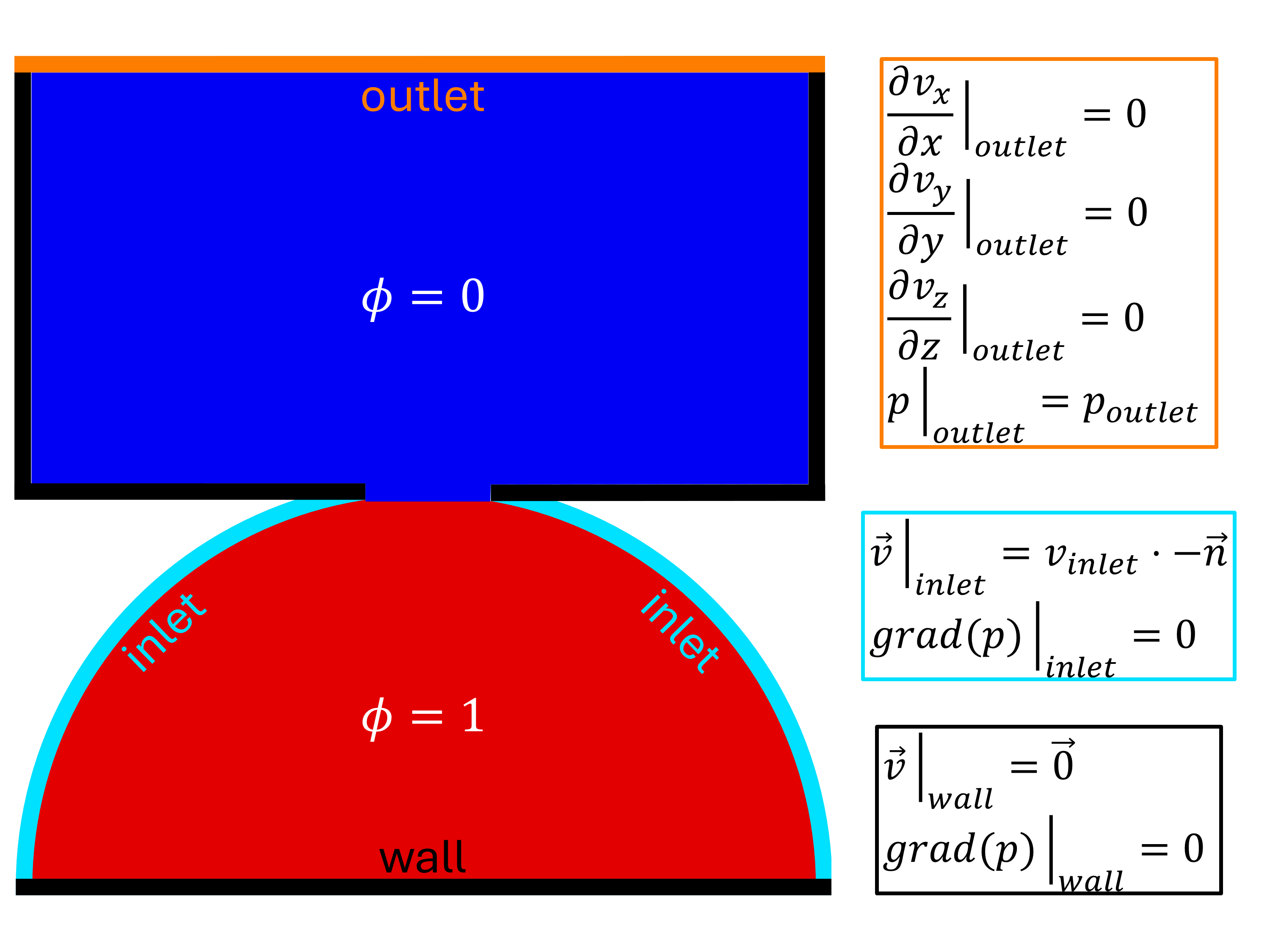}
    \caption{\footnotesize Two-dimensional sketch of the three-dimensional simulation setup and the applied boundary conditions. Red represents the initial lithium phase $\phi =1$ and blue the initial electrolyte phase $\phi =0$. The connection between red and blue represents a crack in the SEI where lithium can flow into the electrolyte. The turquoise inlet represents the intact SEI through which lithium is continuously electroplated. The orange outlet allows for outflow of electrolyte which is pushed away by the extruding lithium. The simulation setup is restricted by walls, represented by black lines. }
    \label{fig:simulation_setup}
\end{figure}

Thirdly, we investigate whisker growth as an extrusion process. We consider a lithium nucleus with a pre-existing crack in the SEI, as shown in Figure~\ref{fig:simulation_setup}. The round-shaped nucleus serves as a model system, but the presented mechanism also holds for more complicated geometry, \textit{e.g.}, crystal-shaped nuclei, or protrusions of heterogeneous lithium foils. The lithium is modeled as a Herschel-Bulkley shear-thinning fluid. A Herschel-Bulkley fluid is a generalized model of a Non-Newtonian fluid that can flow once the yield stress $\sigma_\text{yield}$ is exceeded. The shear-thinning behavior represents the experimentally observed power-law creep of lithium\cite{Ning2023}. The effective viscosity $\mu_\text{eff}$ depends on the shear rate $\dot{\gamma}$, according to
\begin{equation}
\label{eq:Herschel-Bulkley}
    \mu(\dot{\gamma})_\text{eff}=C \left(  \frac{\dot{\gamma}}{\si{s^{-1}}}\right)^{n-1} + \sigma_\text{yield} \cdot \frac{1-\exp{\left(-m\dot{\gamma}\right)}}{\dot{\gamma}} \ ,
\end{equation}
where the dimensionality constant $C$ and the exponent $n$ describe the shear thinning behavior, and $m$ is a regularization parameter for numerical stability of the model. The viscosity is lowered with increasing shear rate.

We use the FLUID software\cite{FLUID} to solve the Navier-Stokes equations for the velocity $\vec{v}$ and the pressure $p$:
\begin{equation*}
    \begin{aligned}
        &\nabla  \vec{v} =0 \ , \\
        &\rho \left( \frac{\partial \vec{v}}{\partial t} + (\vec{v} \cdot \nabla) \vec{v} \right) = - \nabla p + \nabla \cdot \sigma + \rho g \ .
    \end{aligned}
\end{equation*}

We treat lithium as incompressible, \textit{i.e.}, with constant density $\rho$, with shear-thinning Herschel-Bulkley behavior, as described in Equation~\ref{eq:Herschel-Bulkley}. The constitutive equation for the stress $\sigma$ is
\begin{equation*}
    \sigma = 2\mu_\text{eff} \gamma \ ,
\end{equation*}
where 
\begin{equation*}
    \gamma = 0.5 \left(\nabla \vec{v}+\nabla \vec{v}^\text{T}\right) \ .
\end{equation*}

\begin{figure*}[ht]
\centering
\includegraphics[width=\textwidth]{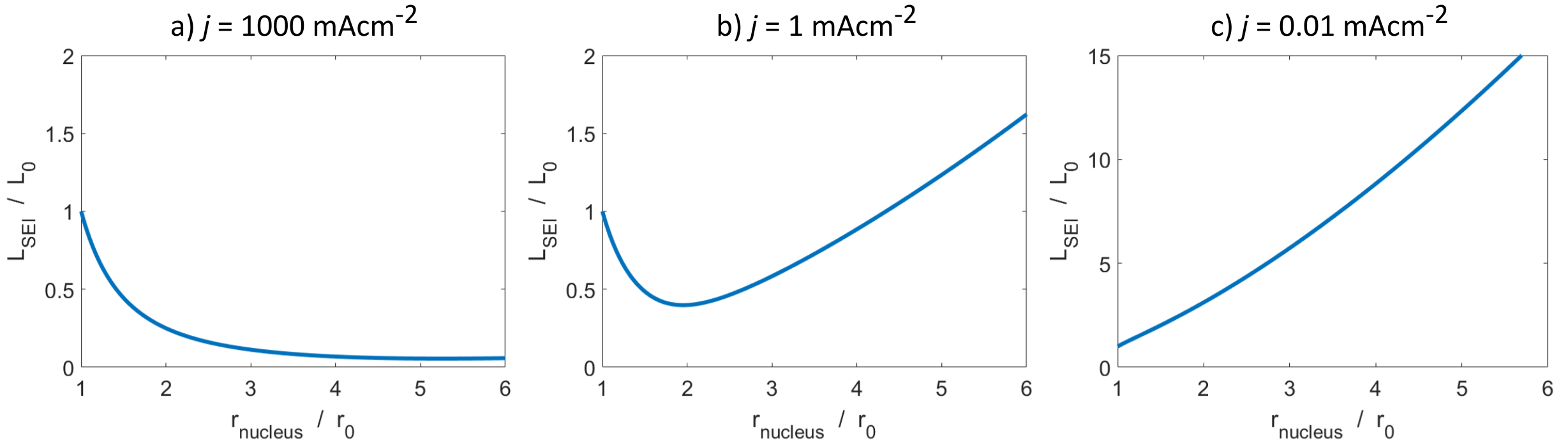}
\caption{ \footnotesize Evolution of SEI thickness $L_\text{SEI}$ covering a growing lithium nucleus with radius $r_\text{nucleus}$ for different current densities, as described by Equation~\ref{eq:SEI_growth}. The initial nucleus radius is $r_0=\SI{1}{\micro \meter}$ and the initial SEI thickness $L_0=\SI{5}{\nano \meter}$. a) High current densities lead to a prolonged thinning of the SEI. b) Intermediate current densities lead to an initial thinning of the SEI which stabilizes quickly. c) Low current densities lead to ongowing SEI growth.}
\label{Fig:SEI_stability}
\end{figure*}

From below, the lithium is bounded by the copper substrate, which is treated as a wall boundary condition. The particle is mostly covered by intact SEI. A steady influx through the intact SEI represents the continuous electrodeposition:
\begin{equation*}
    v_\text{inlet}= \frac{j\cdot V_\text{m}^\text{Li}}{F} \ .
\end{equation*}
The pre-existing crack allows for flow into the electrolyte. We assume that the SEI is rigid enough not to crack further during lithium extrusion. A phase-field model distinguishes between the electrolyte phase $(\phi=0)$ and the Herschel-Bulkley fluid phase $(\phi=1)$. The FLUID software solves the continuity equation,
\begin{equation*}
    \frac{\partial \phi }{\partial t} + \nabla \cdot (\vec{v} \phi) = 0 \ .
\end{equation*}
With this setup, we can simulate the extrusion of lithium.


\section{Results}

To illustrate the relation between current density and SEI thickness, we compare diffusion-limited SEI growth with the SEI deformation due to particle growth. In the case of very high current densities, the fast-growing nucleus stretches the SEI, leading to decreasing SEI thickness, as depicted in Figure~\ref{Fig:SEI_stability} a). The decreasing SEI thickness means the SEI cannot heal fast enough to maintain its passivating property and breaks constantly. In this regime, SEI growth becomes reaction-limited. The nucleus growth outpaces SEI formation, and rhombic dodecahedra-shaped particles form as predicted by the Wulff construction. Boyle \textit{et al.} found that a current density of $\SI{1}{A cm^{-2}}$ leads to Wulff-shaped growth in EC-DEC electrolyte\cite{Boyle2022}. This current density equals a deposition time of $\SI{130}{\micro \second}$ per atom and lattice site. This time scale gives an estimate of the reaction limitation of the SEI formation.

At low current densities, the SEI is estimated to grow continuously despite being stretched by the particle growth, as depicted in Figure~\ref{Fig:SEI_stability} c). Overall, the SEI is stable due to the relatively slow volume expansion, and the self-healing of the SEI prevents the SEI from cracking. The exact value of this regime depends on the self-healing properties of the SEI and its ability to withstand fracture. For example, in localized high-concentration electrolytes, this regime is often achieved at practical current densities in the order of $\SI{1}{mA cm^{-2}}$ and a rounder morphology is observed\cite{Cao2021}. 

The intermediate regime is the regime where the SEI can crack locally due to the high growth stress and cannot heal fast enough. Whiskers can then extrude through these cracks. In our one-dimensional, averaged model, the SEI initially gets thinner but then recovers, as depicted in Figure~\ref{Fig:SEI_stability} b). In carbonate-based electrolytes, this intermediate regime is observed for battery-relevant current densities of $0.1-\SI{100}{mA cm^{-2}}$\cite{Xu2020}.

To summarize, we expect three distinct growth regimes depending on the applied current density and the SEI properties: (1) the covering SEI constantly breaks and does not influence the nucleus growth, (2) the covering SEI does not break, but the deformation of the SEI imposes an additional energy cost to growth, and (3) an intermediate regime, where the SEI locally cracks and whiskers can emerge.

\begin{figure*}[ht]
\centering
    \includegraphics[width=\textwidth]{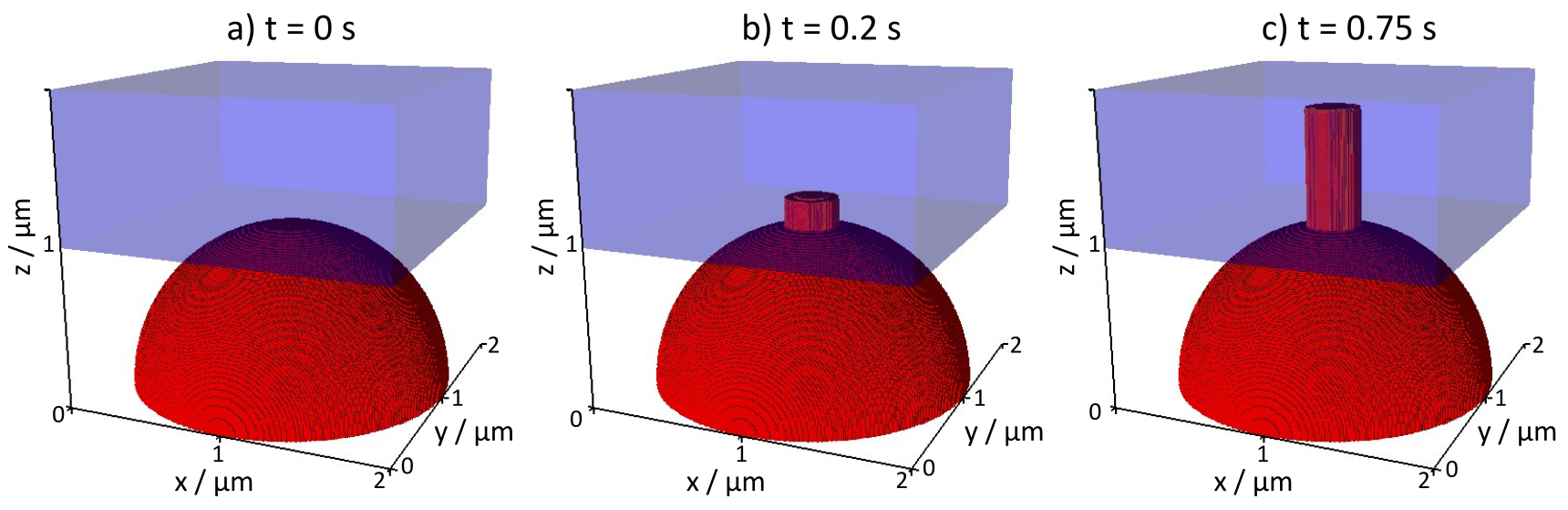}
    \caption{Snapshots of the time evolution of the Herschel-Bulkley fluid phase (red) which is extruded into the electrolyte phase (blue, transparent). a) $t=\SI{0}{s}$ is showing the starting setup. b) after $t=\SI{0.2}{s}$ a protrusion is extruded into the electrolyte phase. c) the extrusion is elongated and is grown in a one-dimensional manner.}
    \label{Fig:time_evolution}
\end{figure*}

\begin{figure*}[ht]
\centering
    \includegraphics[width=\textwidth]{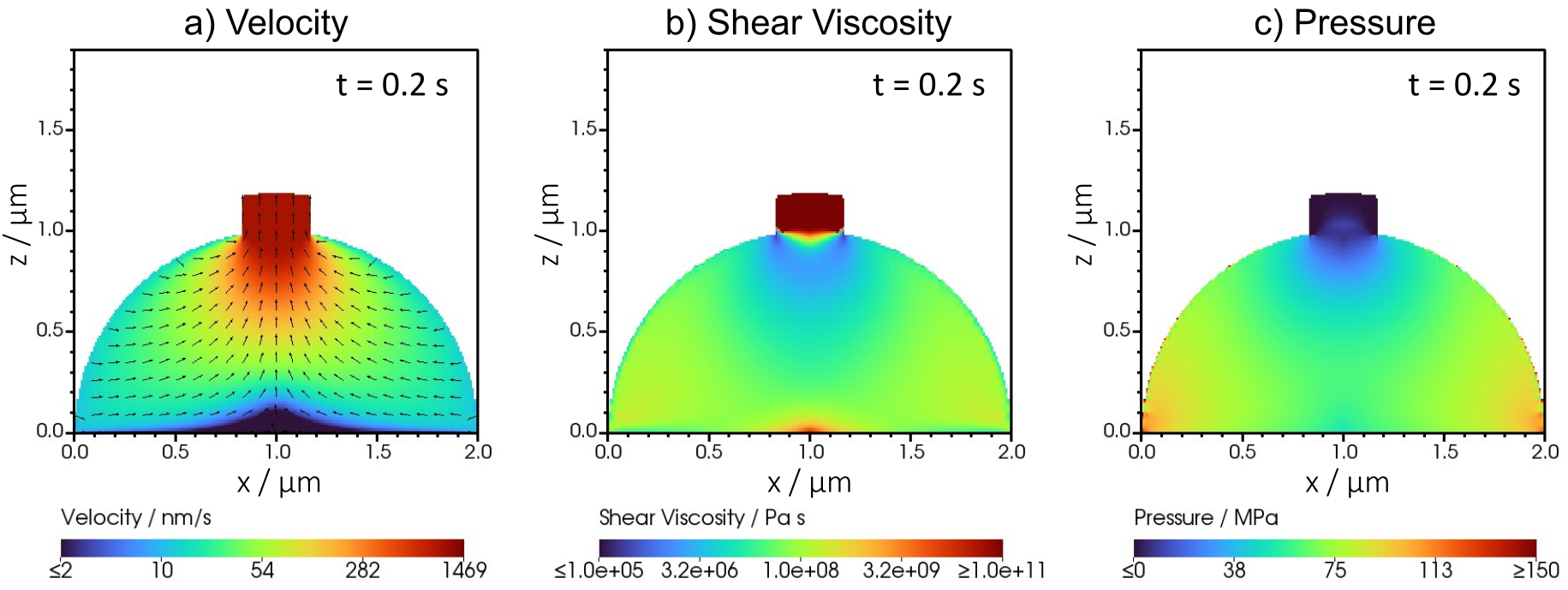}
    \caption{Cross-section of the simulation domain, showing the a) the velocity profile, b) the shear viscosity profile, and c) the pressure profile for the lithium phase ($\Phi \ge 0.9$). a) The magnitude of the velocity is depicted in a logarithmic pseudo-color scheme. The direction of flow is indicated by black arrows. b) The magnitude of the shear-viscosity is depicted in a logarithmic pseudo-color scheme. c) The pressure distribution is depicted in a linear pseudo-color scheme.}
    \label{Fig:variables}
\end{figure*}

\begin{figure}[ht]
\centering
    \includegraphics[width=0.4\textwidth]{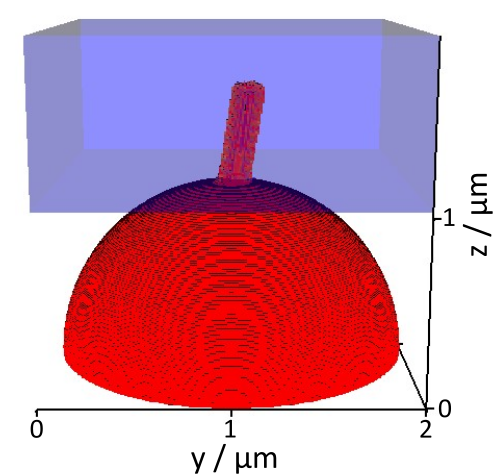}
    \caption{Snapshot of the Herschel-Bulkley fluid phase (red) extruded into the electrolyte phase with a small perturbation near the connection between initial Herschel-Bulkley fluid phase and electrolyte phase. The whisker growth is skewed.}
    \label{Fig:skewed_whisker}
\end{figure}

Based on this estimate, we simulate the extrusion of lithium whiskers in the intermediate regime with $|\Vec{v}_\text{influx}|=\SI{10}{\nano\meter \second^{-1}}\widehat{=}\SI{7.4}{\milli \ampere \centi \meter^{-2}}$. The simulation setup is sketched in Figure~1. We investigate the time evolution of the lithium phase ($\phi\ge0.95$); see Figure~\ref{Fig:time_evolution}. At $t=0$, we show the lithium phase in the starting configuration. At $t=\SI{0.2}{\second}$, a protrusion is grown out of the circular SEI crack. Noticeably, it keeps the circular base area shape. At $t=\SI{0.75}{\second}$, the extruded lithium maintains the cylindrical shape and only grows in length. We interpret this as the typical needle-like whisker growth as observed in experiments. This setup aims to describe the early stage of whisker emergence. We neglect that lithium can electroplate at the whisker and argue that the extrusion of lithium will dominate whisker growth in the early stages. 

To understand the needle-like extrusion, we analyze the velocity field, the shear-viscosity, and the pressure, as depicted in Figure~\ref{Fig:variables}. The flow of the lithium phase, shown in Figure~\ref{Fig:variables} a), is directed toward the crack in the SEI, and plug-like flow is observed in the extruded phase. As all of the inflowing lithium is extruded, the extrusion velocity is two orders of magnitude higher than the influx velocity. This is consistent with the experimental observation that the lithium whiskers grow faster than the applied current density would suggest\cite{Kushima2017}. Thus, extrusion dominates whisker growth in the early stages of whisker emergence. 

The shear-thinning nature leads to lower viscosity in the vicinity of the crack, as depicted in Figure~\ref{Fig:variables} b). This lower viscosity allows for flow towards the crack. It can be seen that the shear viscosity varies over many orders of magnitude in the lithium phase. Once extruded, the lithium phase has a very high shear viscosity. The high shear viscosity resembles the solid-like behavior of lithium once extruded. As it behaves like a solid, it will keep its needle-like shape.  

The driving force for the flow of lithium is a pressure gradient. Figure~\ref{Fig:variables} c) depicts the predicted pressure distribution. The pressure profile shows that pressure is built up inside the initial nucleus. There is a pressure gradient towards the crack in the SEI, and pressure drops off in the extruded phase. The absolute values of the pressure distribution should not be given too much importance as they depend on some parameters that we can only estimate, such as the yield stress. For bulk lithium, the yield stress is reported to be in the range of $ 0.41 - \SI{1.26}{MPa}$\cite{osti_804180,Fincher2020}. For our simulation, we chose a yield stress value of $\sigma_\text{yield}=\SI{10}{\mega \pascal}$, as we investigate small length-scales where yield stress is typically higher than on macroscopic length scales\cite{Fincher2020}. In this case, the predicted pressure inside of the nucleus is in the range of $10-\SI{100}{MPa}$. With a yield stress of $\sigma_\text{yield}=\SI{1}{MPa}$, the predicted pressure is lower, in the order of $\SI{10}{MPa}$; see Supporting Information Figure~SI-2. Estimating the yield stress can be seen as a source of error for a quantitative analysis. However, crucially, the qualitative behavior remains the same. In either case, the SEI will not fracture further. The inorganic SEI components typically have a Young's modulus $E$ in the order of $10-\SI{100}{\giga\pascal}$, and fracture typically occurs for roughly $10\%E \gtrsim \SI{1}{GPa}$. 

Experiments show that whiskers can change growth directions and have kinks\cite{Li2017a,Steiger2014a}. Kushima \textit{et al.} discusses that whisker growth is similar to stick-slip dynamics in friction, where movement can suddenly be interrupted and start again\cite{Kushima2017}. They discuss that kinks can form, possibly when whisker growth suddenly accelerates. There, a new whisker segment grows in a different direction and can push the already existing segments outwards. During the intermittent halt of the whisker growth, a small heterogeneity can form that can induce the quasi-random change in growth direction. We investigate how whisker growth changes by introducing a small perturbation to our setup. We introduce a small inclined wall, which can represent heterogeneous SEI growth. As depicted in Figure~\ref{Fig:skewed_whisker}, this can lead to a change in growth direction. We interpret this accordingly to Kushima \textit{et al.} that heterogeneity can form during intermittent halt of whisker growth, leading to a change in whisker growth directions and inducing kinks. In reality, the angle between two growth directions is probably related to the angle between the crystallographic directions.  
 
\section{Conclusion}

We presented a model for lithium whisker growth as an extrusion of lithium through a crack in the SEI on a lithium nucleus. We modeled the lithium extrusion as the flow of a Herschel-Bulkley shear-thinning fluid. We have shown that this reproduces the experimentally observed needle-like whisker shape. The shear-thinning behavior allows matter to transport to the whisker root. There, lithium is extruded through cracks in the SEI. Once extruded, the lithium behaves solid-like, maintaining the needle-like shape. An everyday analogy would be the shear-thinning behavior of toothpaste\cite{Ahuja2018}. We have shown that small perturbations lead to changes in the whisker growth direction. We interpret the change in growth direction as the formation of a whisker kink. The presented mechanism is studied for a model system of a whisker growing out of a lithium nucleus, as frequently observed when lithium is electroplated on copper\cite{Xu2020}. The same effect can hold for deposition on a lithium foil once planar deposition is no longer possible, and large overpotential leads to nuclei formation, and whiskers then arise as discussed\cite{Rulev2020,Rulev2025}.  

In line with whisker research from other fields, we highlight the stability of the surface covering layer and its role in whisker formation\cite{Ills2017,Majumdar2019}. In the case of lithium electroplating, the SEI is the surface covering layer. A mechanically stable SEI with good self-healing properties can mitigate lithium whisker formation. Based on a rule-of-thumb estimation for carbonate-based electrolytes, we expect unstable SEI growth over a large range of battery-relevant current densities. Given the detrimental effect of lithium whisker formation, the way forward to enable lithium-metal batteries is through electrolyte design for good SEI or introducing an artificial surface layer with sufficiently good properties.

\begin{acknowledgement}

M.W., A.L., and B.H. greatfully acknowledge financial support within the Lillint project (03XP0511A). B.H. and A.L. acknowledge Dominik Kramer for fruitful discussions. 

\end{acknowledgement}

\section{Author's contribution}
M.W.: conceptualization, methodology, software, validation, formal analysis, data curation, writing - original draft, visualization; D.N.: conceptualization, software, validation, investigation, data curation, writing - review \& editing; A.L.: conceptualization, resources, writing - review \& editing, supervision, project administration, funding acquisition; B.H.: conceptualization, methodology, resources, writing - review \& editing, supervision, project administration, funding acquisition.
\begin{suppinfo}

Additional details on the model and experimental findings are provided on-line:
\begin{itemize}
    \item Table of model parameters
    \item Simulation setup and discretization
    \item Impact of yield stress on pressure distribution
\end{itemize}

\end{suppinfo}


{ \scriptsize
\setlength{\parskip}{0pt}
\bibliography{references}
}

\end{document}